\documentclass[12pt,preprint]{aastex}
\usepackage{graphicx}

\begin{document}
\slugcomment{Accepted for Publication in ApJ}

\title{A Deep {\em Chandra} ACIS Study of NGC 4151. I. the X-ray
  Morphology of the 3 kpc-diameter Circum-nuclear Region and Relation
  to the Cold ISM}

\author{Junfeng Wang\altaffilmark{1}, Giuseppina
  Fabbiano\altaffilmark{1}, Guido Risaliti\altaffilmark{1,2}, Martin
  Elvis\altaffilmark{1}, Margarita Karovska\altaffilmark{1}, Andreas
  Zezas\altaffilmark{1,3}, Carole G. Mundell\altaffilmark{4}, Gaelle
  Dumas\altaffilmark{5}, and Eva Schinnerer\altaffilmark{5}}

 \altaffiltext{1}{Harvard-Smithsonian Center for Astrophysics, 60 Garden St, Cambridge, MA 02138}
 \altaffiltext{2}{INAF-Arcetri Observatory, Largo E, Fermi 5, I-50125 Firenze, Italy}
 \altaffiltext{3}{Physics Department, University of Crete, P.O. Box 2208, GR-710 03, Heraklion, Crete, Greece}
 \altaffiltext{4}{Astrophysics Research Institute, Liverpool John Moores University, Birkenhead CH41 1LD, UK}
 \altaffiltext{5}{Max-Planck-Institut f$\ddot{\rm u}$r Astronomie, K$\ddot{\rm o}$nigstuhl 17, D-69117 Heidelberg, Germany}

\email{juwang@cfa.harvard.edu}

\begin{abstract}

We report on the imaging analysis of $\sim$200 ks sub-arcsecond
resolution \emph{Chandra} ACIS-S observations of the nearby Seyfert 1
galaxy NGC 4151.  Bright, structured soft X-ray emission is observed
to extend from 30 pc to 1.3 kpc in the south-west from the nucleus,
much farther than seen in earlier X-ray studies.  The terminus of the
north-eastern X-ray emission is spatially coincident with a CO gas
lane, where the outflow likely encounters dense gas in the host
galactic disk.  X-ray emission is also detected outside the boundaries
of the ionization cone, which indicates that the gas there is not
completely shielded from the nuclear continuum, as would be the case
for a molecular torus collimating the bicone.  In the central
$r<200$~pc region, the subpixel processing of the ACIS data recovers
the morphological details on scales of $<$30~pc ($<$0.5\arcsec) first
discovered in {\em Chandra} HRC images.

The X-ray emission is more absorbed towards the boundaries of the
ionization cone, as well as perpendicular to the bicone along the
direction of a putative torus in NGC 4151.  The innermost region where
X-ray emission shows the highest hardness ratio, is spatially
coincident with the near-infrared resolved H$_2$ emission and dusty
spirals we find in an $HST$ $V-H$ color image.  The agreement between
the observed H$_2$ line flux and the value predicted from
X-ray-irradiated molecular cloud models supports photo-excitation by
X-rays from the active nucleus as the origin of the H$_2$ line,
although contribution from UV fluorescence or collisional excitation
cannot be fully ruled out with current data.  The discrepancy between
the mass of cold molecular gas inferred from recent CO and
near-infrared H$_2$ observations may be explained by the anomalous CO
abundance in this X-ray dominated region.  The total H$_2$ mass
derived from the X-ray observation agrees with the measurement in
\citet{SB09}.

\end{abstract}


\section{Introduction}

NGC 4151 \citep[classified as (R$^{'}$)SAB(rs)ab;][]{deVaucouleurs91},
one of the nearest ($d\sim 13.3$ Mpc, $1\arcsec=65$ pc; Mundell et
al. 1999), and apparently brightest, active galaxy (Seyfert 1.5,
Osterbrock \& Koski 1976), has been extensively studied in almost all
wavebands (see Ulrich 2000 for a review). Because of its proximity,
NGC 4151 offers a unique opportunity to examine the fueling of the
central supermassive black hole (SMBH) and its impact on the host
galaxy.

NGC 4151 is close to face-on (inclination $i \approx 21^{\circ}$) with
a major axis position angle (P.A.) $\approx 22^{\circ}$ (Pedlar et
al. 1992; Mundell et al. 1999).  A biconical extended narrow line
region (ENLR) is elongated along the northeast (NE) and southwest (SW)
of the active galactic nucleus (AGN), at P.A. $\sim$65$^{\circ}$ and
230$^{\circ}$, respectively (Perez et al.\ 1989; P{\'e}rez-Fournon \&
Wilson 1990; Penston et al. 1990; Evans et al. 1993).  Clumpy ionized
gas is seen in {\it Hubble Space Telescope} ({\em HST}) narrow-band
images (e.g., [OIII] $\lambda$5007\AA\/, Boksenberg et al.\ 1995; Winge et
al.\ 1999) outflowing along the bicone (Schulz 1990; Crenshaw et
al.\ 2000; Kaiser et al.\ 2000; Das et al.\ 2005), and extending to
$\sim$1 kpc away from the AGN.  The morphology, kinematics, and line
ratio diagnostics from optical studies of NGC 4151's ENLR (e.g.,
Robinson et al. 1994; Hutchings et al. 1998; Das et al. 2005) strongly
support that AGN radiation in the optical bicone intersects with the
host galactic disk, and photoionizes ambient galactic gas that is
participating in normal galactic rotation
\citep{Pedlar92,Evans93,Asif98,Mundell99}.

Near-infrared (NIR) emission line mapping of the narrow line region
(NLR) of NGC 4151 \citep[e.g., Pa$\beta$;][]{SB09,SB10} shows that the
ionized gas closely follows the optical bicone, and extends outside
the bicone (also found in the HST study by Kraemer et al.\ 2008).  The
molecular hydrogen (H$_2$) emission distribution is perpendicular to
the bicone axis \citep{Fer99,SB09}, following approximately the
P.A. of a large scale ($3.2\arcmin \times 2.5\arcmin$) gas-rich
stellar bar \citep[$\sim$130$^{\circ}$;][]{Pedlar92,Mundell99}.
\citet{SB10} propose that H$_2$ emission is potentially excited by
X-rays from the AGN or shocks related to the accretion flow along the
large scale bar.  A spatially resolved study of the $^{12}$CO emission
\citep{Dumas10} finds two prominent CO gas lanes at 1 kpc from the
nucleus, but no cold molecular gas in the central 300 pc where the NIR
H$_2$ line emission peaks. \citet{Dumas10} further suggest that the
H$_2$ line emission is likely photo-excited by the AGN.

The ENLR of NGC 4151 has been resolved in the soft X-rays with
$Einstein$ and $ROSAT$ observations (Elvis, Briel, \& Henry 1983;
Morse et al. 1995).  Early {\em Chandra} images show that the extended
soft X-ray emission is generally correlated with the optical forbidden
line emission (e.g., Ogle et al. 2000; Yang et al. 2001), whereas the
nucleus remains unresolved in the hard X-ray band.  Our recent deep
{\em Chandra} observations of NGC 4151 (PI: Fabbiano) on the
circum-nuclear region are now providing detailed information on
several spatial scales.  The {\em Chandra} HRC image (Wang et
al. 2009) resolves the nuclear X-ray emission on spatial scales of
$\sim$30 pc, finding evidence for interactions between the radio
outflow and NLR clouds\footnote{For a comprehensive view of the NGC
  4151 nuclear structure, see Mundell et al. (2003).}.  Our deep ACIS
data have led to the discovery of a large scale soft diffuse X-ray
emission (Wang et al. 2010a) extending out to $\sim$2 kpc from the
active nucleus and filling in the cavity in the HI 21 cm emission
distribution \citep{Asif98,MS99}.  This extended X-ray emission could
be due to either hot gas in the galactic disk heated by the nuclear
outflow or gas photoionized by an outburst of the AGN (Komossa 2001;
Wang et al. 2010a).

This is the first of a series of papers where we report on the results
from deep {\em Chandra} ACIS observations of the circum-nuclear region
in NGC 4151, focusing on the imaging analysis and relating the X-ray
morphology to the cold interstellar medium (ISM) in the host galactic
disk.  This paper is organized as follows. In \S~2 we describe the
observations and data reduction; in \S~3 we describe the imaging
analysis and present the X-ray morphology.  In \S~4 we discuss the
results, in particular the role of the X-ray emission in exciting the
molecular material near the AGN.  We summarize our findings in \S~5.
In a forthcoming paper (Wang et al. in preparation; Paper II), we will
present the X-ray emission line strength maps, detailed comparison of
these maps with the ionized gas distribution using H$\alpha$ and
[OIII] images, and spatially resolved spectral fitting with
self-consistent photoionization models.

\section{Observations and Reduction}

\subsection{$Chandra$ data}\label{obs_1}

NGC 4151 was observed with {\em Chandra} for a total of $\sim$200 ks,
using the spectroscopic array of the Advanced CCD Imaging Spectrometer
(ACIS-S; Garmire et al. 2003) in 1/8 sub-array mode during March
27-29, 2008.  The nucleus of NGC 4151 was placed near the aimpoint on
the backside-illuminated S3 chip, which has a high sensitivity to soft
X-rays.  The data were analyzed following the standard procedures
using CIAO\footnote{\url{http://cxc.harvard.edu/ciao/}} (Version 4.2)
with CALDB\footnote{\url{http://cxc.harvard.edu/caldb/}} (Version
4.2.1) provided by the {\em Chandra} X-ray Center (CXC). The
lightcurve from regions free of bright sources on the S3 chip was used
to remove times of high background count rates.  The cleaned data have
total good exposure times of 180 ks (116 ks and 63 ks, in ObsID 9217
and ObsID 9218, respectively).  Table~1 summarizes the observations
used in this paper.

On the 1/8 S3 chip, a total of 24 point sources were detected using a
significance threshold $10^{-6}$ with the wavelet-based algorithm {\sc
  wavdetect\/} (Freeman 2002), part of the CIAO tools.  The search was
performed in wavelet scales from 1 to 16 pixel (0.5\arcsec\/ to
8\arcsec) in steps of $\sqrt{2}$.  These sources are removed from the
images when we perform the spectral extraction of the diffuse
emission.  In addition, matching the positions of the X-ray detections
to the $SDSS$ Data Release 7 catalog indicates an absolute astrometric
accuracy of $0.3\arcsec$.

The two ACIS observations of NGC 4151 were taken almost uninterrupted,
with identical roll angles. The events were reprojected to the
reference frame of ObsID 9217, and merged to create a single image.
The pipeline randomization was removed with the CIAO tool
$acis\_process\_events$.  Subpixel event repositioning and binning
techniques (Tsunemi et al. 2001; Mori et al. 2001; Kastner et
al. 2002; Li et al. 2003, 2004) were applied to improve the effective
resolution of the ACIS images to better than $0.4\arcsec$ (Li et
al. 2004).  The images with subpixel resolution presented in this
paper used the energy-independent shift for the corner split and
2-pixel split events (``static'' method in Li et al. 2004).

Spectra and instrument responses were generated using CIAO Version
4.2. The source spectra were analyzed with XSPEC Version 12.5 (Arnaud
1996), taking background spectra from a blank region from the same CCD
node outside of the galaxy.  Spectra were grouped to have a minimum of
20 counts per energy bin to allow for $\chi^2$ fitting.  When
extracting spectra, the BACKSCAL keyword (equal to the fractional area
of the chip occupied by the extraction region) was computed for the
extended regions using CIAO tool {\em
  specextract}\footnote{\url{http://cxc.harvard.edu/ciao/threads/specextract/}}.
We restricted our modeling to photon energies between 0.3 keV and 7
keV, to avoid the instrument calibration uncertainty at lower
energies, and the higher background at high energies.  The 90\%
confidence interval for a single interesting parameter is reported for
all spectral fitting results.

\subsection{Archival Optical/NIR imaging}\label{obs_2}

To examine extinction with a color map of the circum-nuclear region of
NGC 4151, we have also retrieved from the Hubble Legacy
Archive\footnote{\url{http://hla.stsci.edu}} the $HST$ ACS-HRC/F550M
($V$-band) and NICMOS-NIC2/F160W ($H$-band) images.  These images were
acquired as part of the Cycle 12 Program 9851 (PI: Peterson) and Cycle
7 Program 7215 (PI: Thompson), and the details of these observations
were published in \citet{Onken07} and \citet{Thompson99},
respectively. Pipeline processed images are further cleaned to remove
cosmic rays using the Laplacian Cosmic Ray Identification algorithm
(van Dokkum 2001).  Model PSFs were generated with the TinyTim tool
(Krist 1995) for the specific instrument and filter, and subtracted to
remove the unresolved nuclear component.  The resolution of the ACS
$V$-band image has been degraded to match that of the $H$-band, by
convolving with the broader NICMOS PSF ($FWHM=0.16\arcsec$).

\section{Imaging Analysis and Results}

\subsection{The ``Raw'' ACIS Image}

\begin{figure}
\epsscale{1.0} \plotone{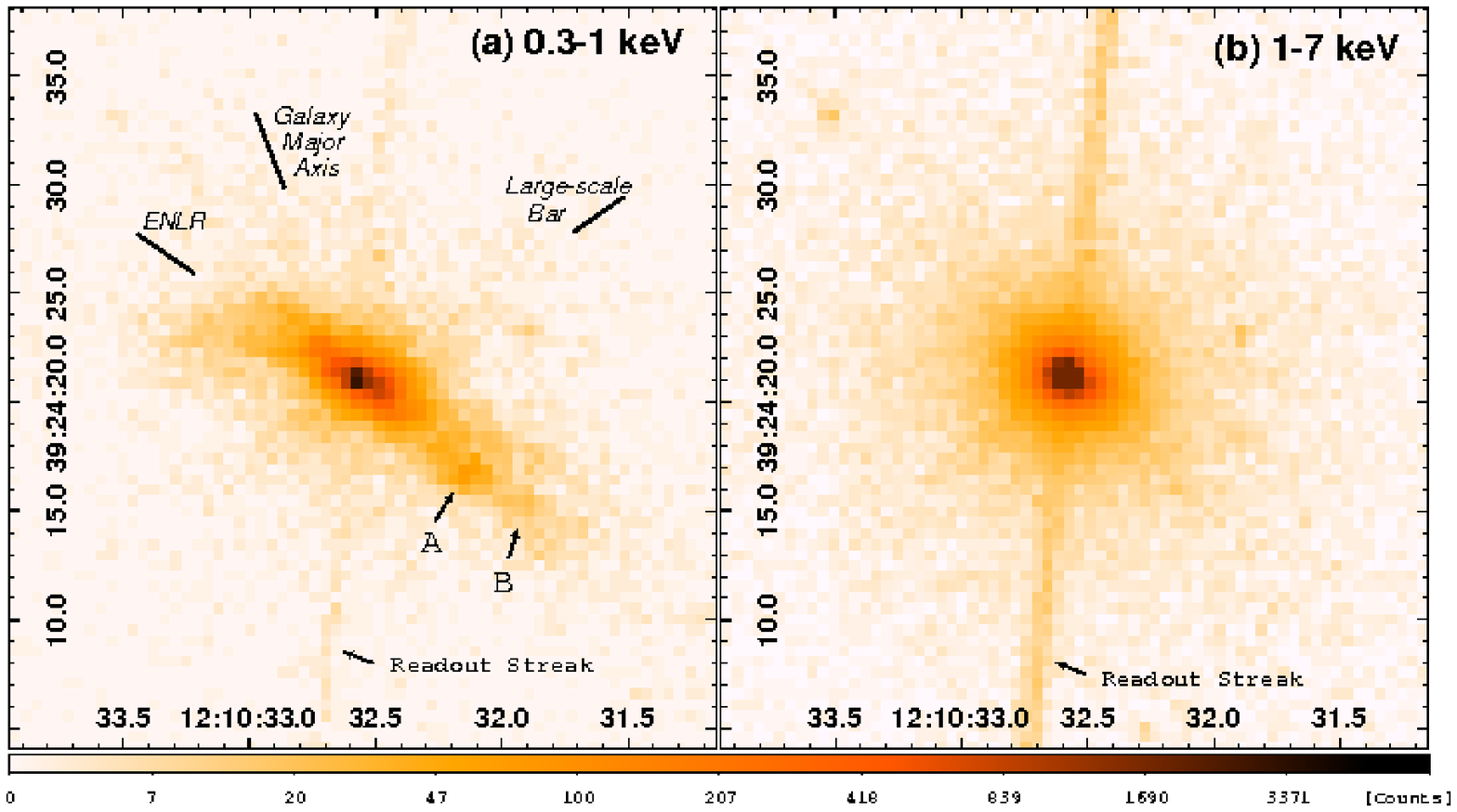} \caption{ACIS raw image of the NGC
  4151 nuclear region (a) in the soft band (0.3--1 keV). (b) the hard
  band (1--7 keV). X-ray enhancements ``A'' and ``B'' are indicated
  with arrows.  Solid lines mark the directions of the kinematic major
  axis of the host galaxy (P.A.$\sim$22$^{\circ}$, Pedlar et al. 1992;
  Mundell et al. 1999), the large scale ``weak fat bar''
  (P.A.$\sim$130$^{\circ}$; Mundell \& Shone 1999), and the ENLR
  bicone (P.A.$\sim$65$^{\circ}$, Evans et al. 1993). The narrow
  streak running north-south across the nucleus (present in both
  panels) is an artifact of the CCD readout of the bright
  nucleus.\label{fig1}}
\end{figure}

\emph{Chandra} ACIS images of the central $\sim 30\arcsec \times
30\arcsec$ ($\sim 2 \times 2$~kpc) region of NGC 4151 were extracted
in the 0.3--1 keV and 1--7 keV energy bands based on its X-ray
spectrum.  These images, shown in Figure~\ref{fig1}, are in good
agreement with the ACIS-S image obtained by Yang et al. (2001) and the
HETG zero-order image shown by Ogle et al. (2000). However, our deeper
images are $\sim$7 and 20 times more sensitive than the previously
published ACIS images of this circum-nuclear region, respectively.
Figure~\ref{fig1} clearly shows the bright nucleus dominating at
energies $>$1~keV and soft resolved extended regions (the streaks on
both sides of the nucleus are an artifact of the CCD readout process).

The SW side (along P.A.$\sim$233$^{\circ}$) of the 0.3--1 keV emission
is clumpy and clearly seen to $r=14\arcsec$, with fainter diffuse
emission extending as far as 20\arcsec\/ from the nucleus. A
relatively bright ``blob'' to the SW, noted in Yang et al. (2001), is
located at $r=6\arcsec$ (380 pc) from the nucleus at a P.A. of
$\sim$227$^{\circ}$ (Figure~\ref{fig1}; indicated as ``A'' with an
arrow). Our image also shows at least one more such clump of X-ray
enhancement further out at 10\arcsec\/ from the nucleus (``B'' in
Figure~\ref{fig1}).  On the NE side, the soft emission is brightest at
$r<5\arcsec\/$ and extends to 11\arcsec\/ from the nucleus along
P.A.$\sim$75$^{\circ}$.  The soft X-ray emission along the NW--SE
directions is fainter than the NE-SW emission by a factor of 2 and is
less extended (3\arcsec; $\sim$200 pc).

\subsection{The Nucleus: Photon pile-up and the Nuclear Point Spread Function (PSF)}\label{pile-up}

The {\em Chandra}--measured position of the Seyfert X-ray nucleus is
$\alpha_{\rm x}(2000) = 12^{\rm h} 10^{\rm m} 32.59^{\rm s}$,
$\delta_{\rm x}(2000) = +39^{\circ} 24^{\prime} 21.2^{\prime\prime}$.
Given the uncertainty in the absolute astrometry, this position is
identical to the optical ($\alpha=12^h 10^m32.^s6$,
$\delta=+39^{\circ}24^{\prime}21^{\prime\prime}$, Clements 1981) and
radio nuclear positions ($\alpha=12^h 10^m32.^s6$,
$\delta=+39^{\circ}24^{\prime}21^{\prime\prime}$, Ulvestad et
al.\ 2005).

The nucleus of NGC 4151 is a well known strong X-ray source.  During
the exposure of a single ACIS frame, two or more photons from such a
high count rate X-ray source may be recorded as a single event.  This
is known as the ``pile-up effect''\footnote{For more information see {\em
    Chandra} ABC Guide to Pile Up, available at
  \url{http://cxc.harvard.edu/ciao/download/doc/pileup\_abc.ps}},
which causes the loss of information from the original events,
resulting in a hardened source spectrum and grade migration (Davis
2001).  Aiming to study the extended X-ray emission as close to the
nucleus as possible, our observations were performed using the ACIS
``1/8 sub-array'' mode (0.6 s frame time) to reduce nuclear pile-up.
However, pile-up is still a concern as Yang et al. (2001) showed that,
although their observation was taken when NGC 4151 was experiencing a
low flux state, pile-up still significantly affects the inner
$1.3\arcsec$-radius region in the ACIS image in the same mode.  Before
we proceed to further imaging analysis, we must evaluate the effects
of pile-up on the nuclear PSF, to exclude that some of the observed
features may be due to pile-up effects.

We first compared the radial profiles of the observed X-ray emission
to that expected from a point source.  To simulate the nuclear PSF, we
need to know the intrinsic spectrum of the heavily piled-up nucleus.
The events extracted from the ACIS readout streak do not suffer
pile-up as the incoming photons during each frame readout were still
recorded by the ACIS CCD with an exposure equivalent to 40 $\mu$s per
frame.  We used these readout events to model the spectrum of the
nucleus (Wang et al. 2010b) and obtained an absorption corrected
luminosity of $L_{2-10\rm{keV}}=1.3\times 10^{42}$ erg s$^{-1}$.

This model was used as the input spectrum to the {\em Chandra}
HRMA\footnote{High Resolution Mirror Assembly (van Speybroeck et
  al.\ 1997); see
  {\url{http://cxc.harvard.edu/cal/Hrma/users\_guide/}} for details.}
ray tracer (ChaRT\footnote{\url{http://cxc.harvard.edu/chart/}};
Carter et al.\ 2003) which was run with the same observation
configurations as our ACIS observations (Table~1).  The ChaRT output
was then fed into the MARX
simulator\footnote{\url{http://space.mit.edu/cxc/marx/}} to project
rays onto the detector and create an event file, with responses
applied.  We also used MARX to simulate a piled event file using the
$\sim$35\% pile-up fraction found by Wang et al. (2010b), in order to
compare with the observed radial profiles.  The effectiveness of this
procedure was demonstrated by Russell et al. (2010).

\begin{figure}
\epsscale{1.2}
\plottwo{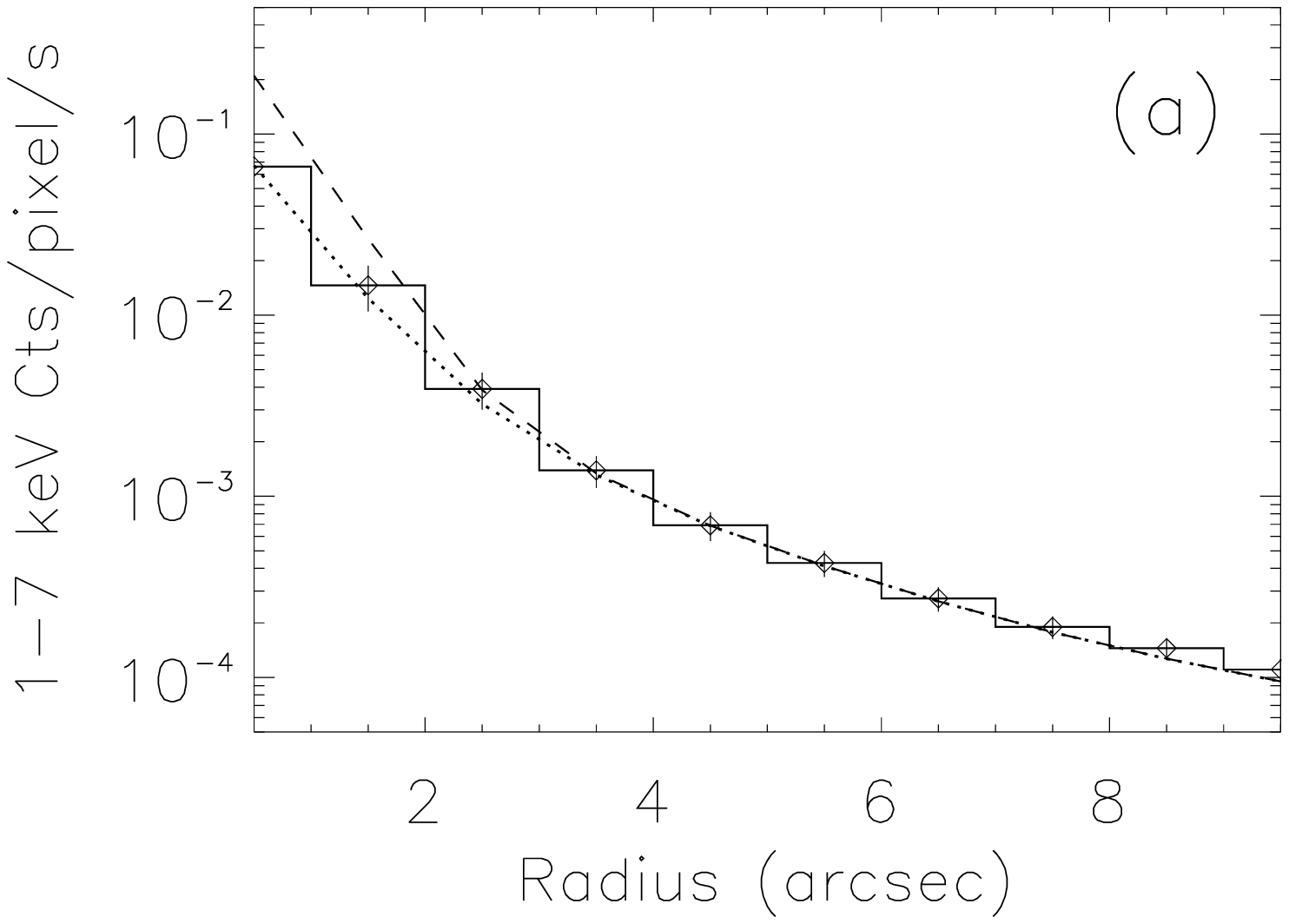}{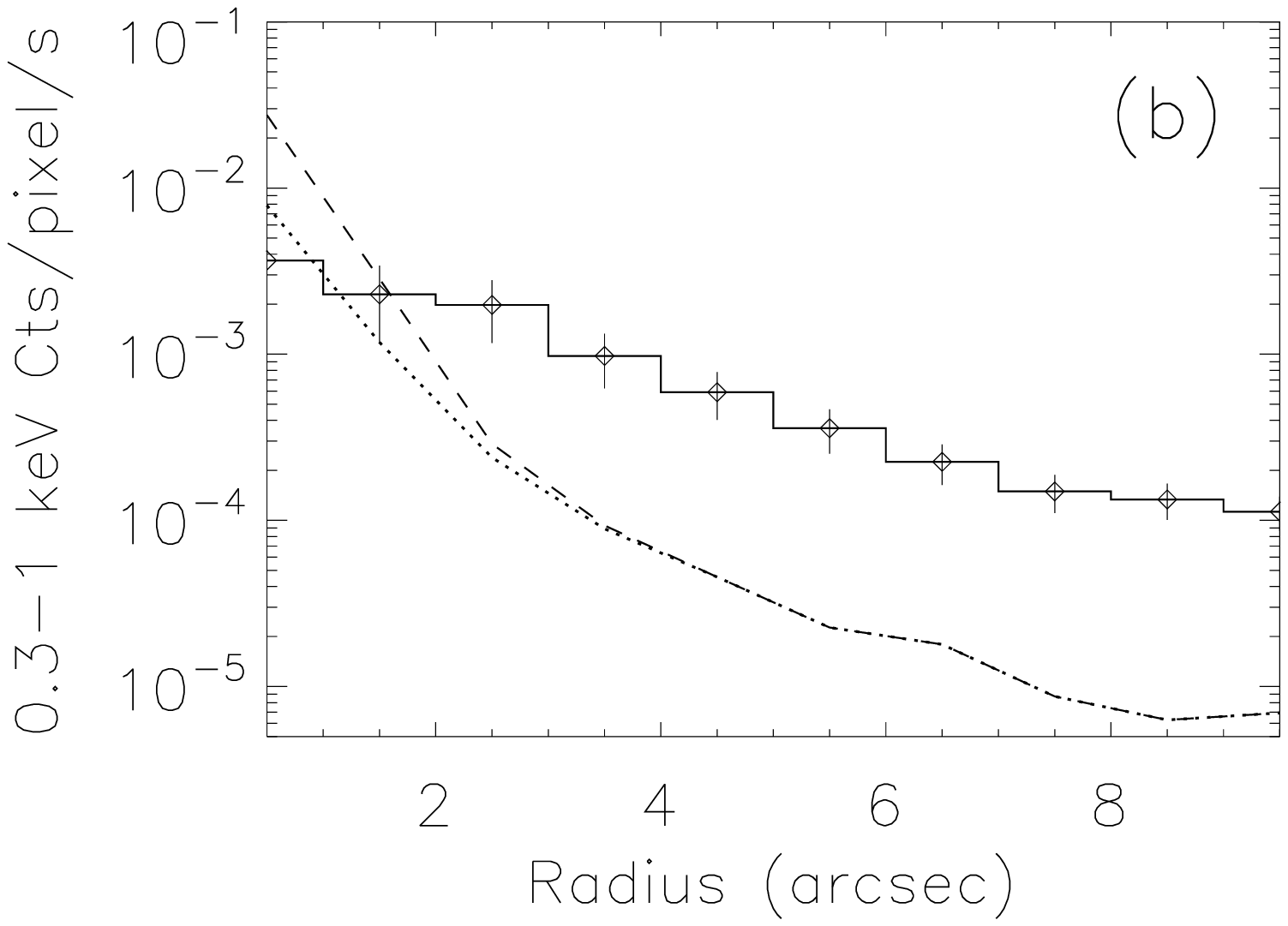}
\plottwo{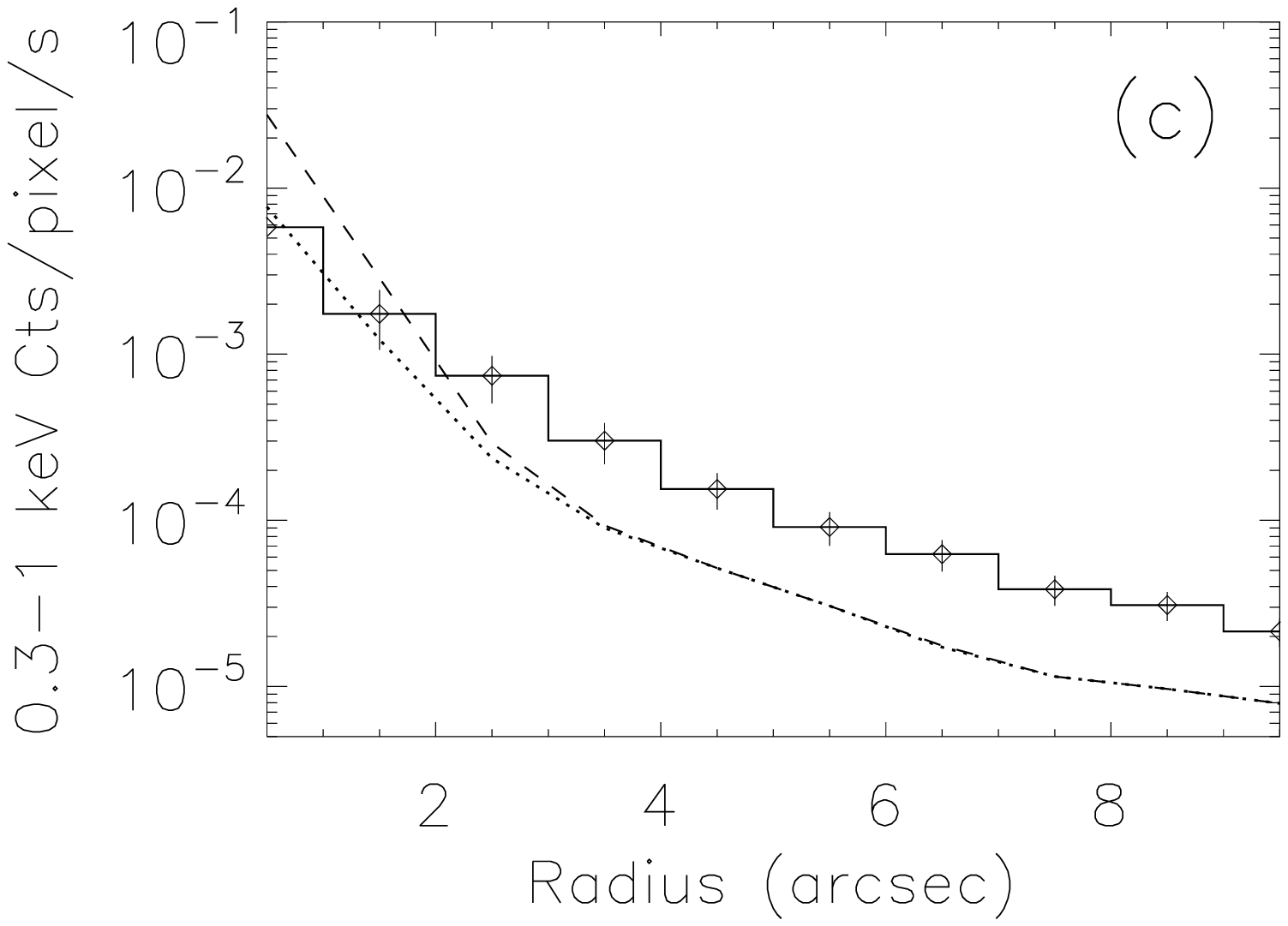}{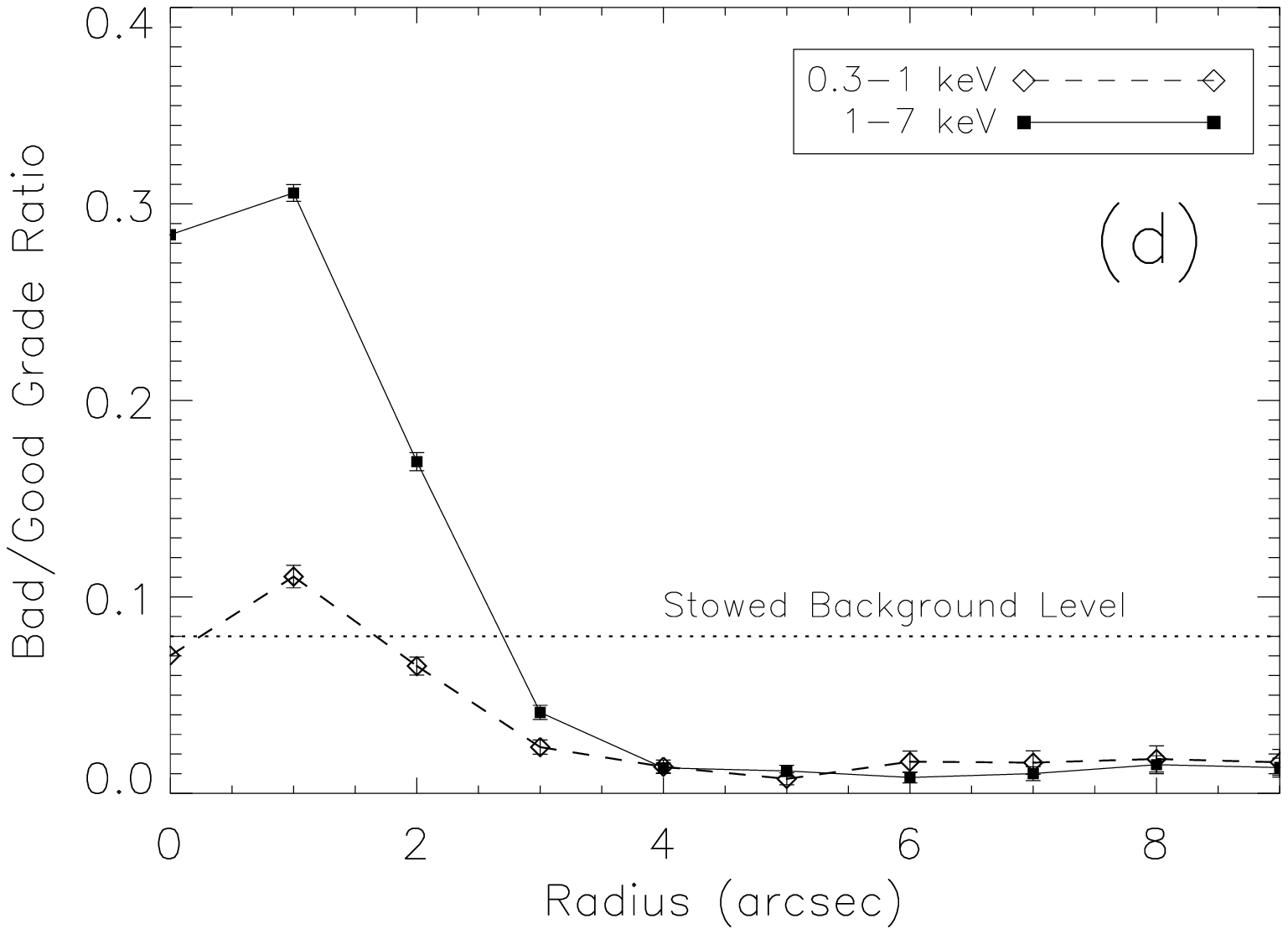}
\caption{(a) Comparison of model PSFs (dashed line: without pile-up
  effect; dotted line: pile-up included) with the observed radial
  profile of the central bright point-like source in the hard band
  (1--7 keV). (b) Comparison of model PSFs with the observed radial
  profile in the 0.3--1 keV band for the obviously extended NE-SW
  sector. (c) the same as (b) but for the less clearly extended SE-NW
  sector. (d) Radial variation for the ratio between events with bad
  grades (1,5,7) and those with good grades (0,2,3,4,6).  The level of
  such ratio in the {\em Chandra} ACIS-stowed background (pile-up
  free) is shown (dotted line). Both the 0.3--1 keV and 1--7 keV bands
  are shown.
\label{fig2}}
\end{figure}

Figure~\ref{fig2}a compares the observed radial profiles of the
nuclear region of NGC 4151 in the 1--7 keV band with the simulated PSF
of a point-like nucleus (with and without pile-up).  The count rate in
the brightest inner 2 arcsec region is significantly lower than the
expected values without pile-up, and consistent with the PSF including
MARX simulated pile-up.  The soft (0.3--1 keV) X-ray emission close to
the nucleus is clearly resolved, both along the NE -- SW and the NW --
SE direction.  In both directions, the soft extended emission
dominates over the nuclear PSF wings at radii $>2\arcsec$ from the
nucleus (Figure~\ref{fig2}b and~\ref{fig2}c).  Beyond a radius of
$\sim$3\arcsec\/ from the nucleus, the piled MARX simulations produced
essentially identical results to the simulations without pile-up, so
at these larger radii pile-up is not an issue.

As pile-up combines events of different energies and causes grade
migration, another way to estimate the importance of pile-up is using
the ratio of events with ``bad'' ASCA grades (1, 5, 7) to those with
``good'' ASCA grades (0, 2, 3, 4, 6) where a bad/good grade ratio of
$<0.1$ indicates that pile-up is approximately $<10\%$ (Russell et
al. 2010).  The radial distribution of the bad/good grade ratio is
shown in Figure~\ref{fig2}d for both the 0.3--1 keV and the 1--7 keV
bands, which indicates that, beyond $r\sim 2\arcsec$ from the nucleus
pile-up is no longer significant in the 0.3--1 keV band (and beyond
$r\sim 3\arcsec$ for the 1--7 keV band).  We conclude that for our
analysis of the {\em extended soft} X-ray emission, the effect of
pile-up is mild beyond a radius of 2\arcsec\/ from the nucleus, and
effectively absent at $r\ga 3\arcsec$.

For imaging analysis, the major effect of pile-up on the ACIS on-axis
PSF is that, the core (where most of the photons hit) gets suppressed,
because some piled events are rejected as bad events, creating a
circular ``caldera'' (e.g., Yang et al. 2001).  At the pile-up
fraction of $\sim$35\% encountered in this observation, there is
$\sim$40\% decrease in brightness in the core of the PSF, but little
distortion of its circular shape (Gaetz et al., private communication;
see the {\it Chandra Proposers' Observatory Guide}
[POG]\footnote{Available at
  \url{http://cxc.harvard.edu/proposer/POG/}} Section 6.15.).  Thus,
although we will discuss the images of extended features in the
central 2\arcsec\/ region that are affected by pile-up with caution,
deviations from circular shape must be related to real X-ray emission
features.

\subsection{ACIS Images with Subpixel Binning}\label{sec:subpix}

{\em Chandra} HRMA's PSF (half-power diameter HPD=0.6\arcsec\/, $E\leq
1.5$~keV; see the {\em Chandra} POG) is undersampled by the ACIS CCD
pixels ($0.492\arcsec\times 0.492\arcsec$).  To take advantage of the
telescope dithering\footnote{see POG Version 12 \S~5.8.2, page 75
  (2009).} to improve the sampling, a fine pixel size
(0.0625\arcsec\/, $\sim$1/8 of the native ACIS pixel size) was used
when extracting the images.  This subpixel binning approach is
frequently adopted in imaging studies of X-ray jets pushing for the
highest spatial resolution (e.g., Harris et al. 2004; Siemiginowska et
al.  2007; Perlman et al. 2010).  In this section we will also make
use of the 50 ks {\em Chandra} observation of NGC 4151 with the High
Resolution Camera (HRC; Murray et al. 1997) presented in Wang et
al.\ (2009).

\begin{figure}
\epsscale{1.0} \plotone{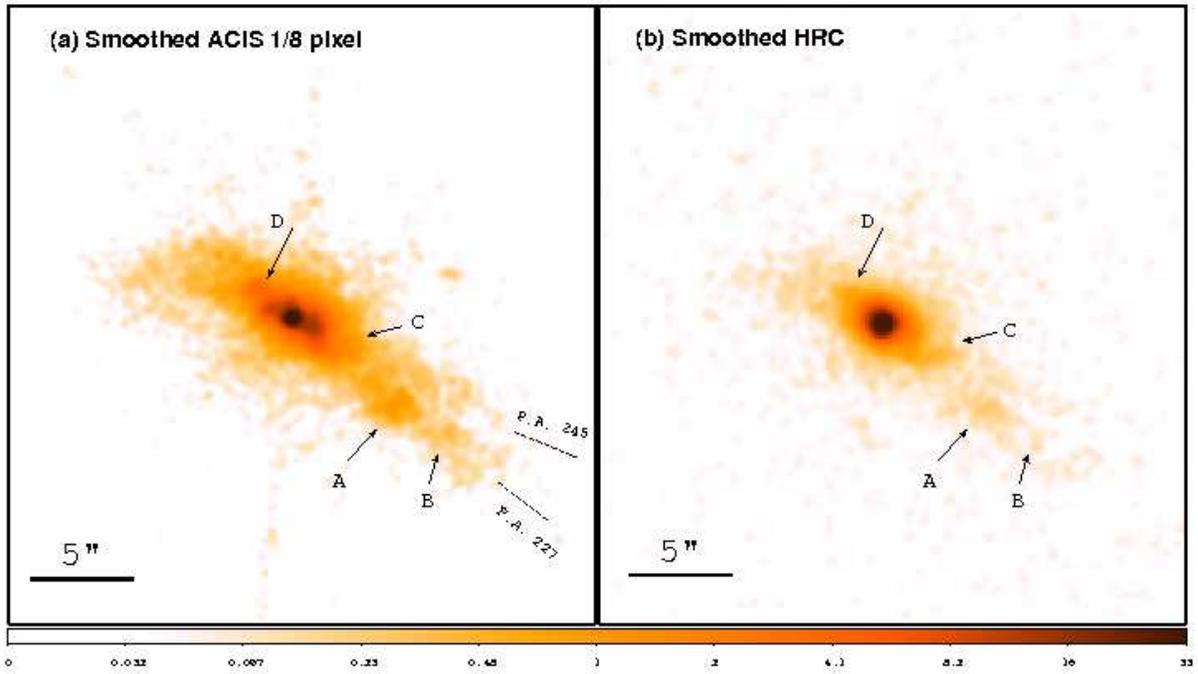} \caption{(a) 0.3--1 keV ACIS
  subpixel image of NGC 4151 (1/8 native pixel); (b) 0.1--10 keV HRC
  image.  Both are smoothed by a $FWHM=0.4\arcsec$ Gaussian kernel for
  better visualization of faint extended features.\label{fig3}}
\end{figure}

Figure~\ref{fig3} compares the 0.3--1 keV subpixel ACIS and HRC (Wang
et al.\ 2009) images both smoothed with a $FWHM=0.4\arcsec$ gaussian
kernel.  Although the raw soft band ACIS image (Figure~\ref{fig2}a)
already shows the presence of extended emission along the NE--SW
direction and hints of split emission in both NE and SW cones, the
subpixel processed ACIS image of Figure~\ref{fig3}a reveals more
morphological details, especially the presence of several X-ray blobs
(marked as ``A'', ``B'', ``C'', and ``D'' in Figure~\ref{fig3}) and
the clear bifurcation of both the NE and SW cones.  The appearance of
the NE--SW soft X-ray emission is an elongated ``X''-shape with an
acute apex angle, resembling edge-brightened cones
(c.f. Figure~\ref{fig1}a). We have measured the P.A. of the brighter
bifurcated SW X-ray emission: the edges are located at $227^{\circ}\pm
8^{\circ}$ and $245^{\circ}\pm 5^{\circ}$, with a full opening angle
of $\sim 20^{\circ}$.  The HRC image shows the identical extended
morphology for the circum-nuclear region (Figure~\ref{fig3}b),
validating that subpixel repositioning does not introduce unwanted
artifacts.

\begin{figure}
\epsscale{1.0} \plotone{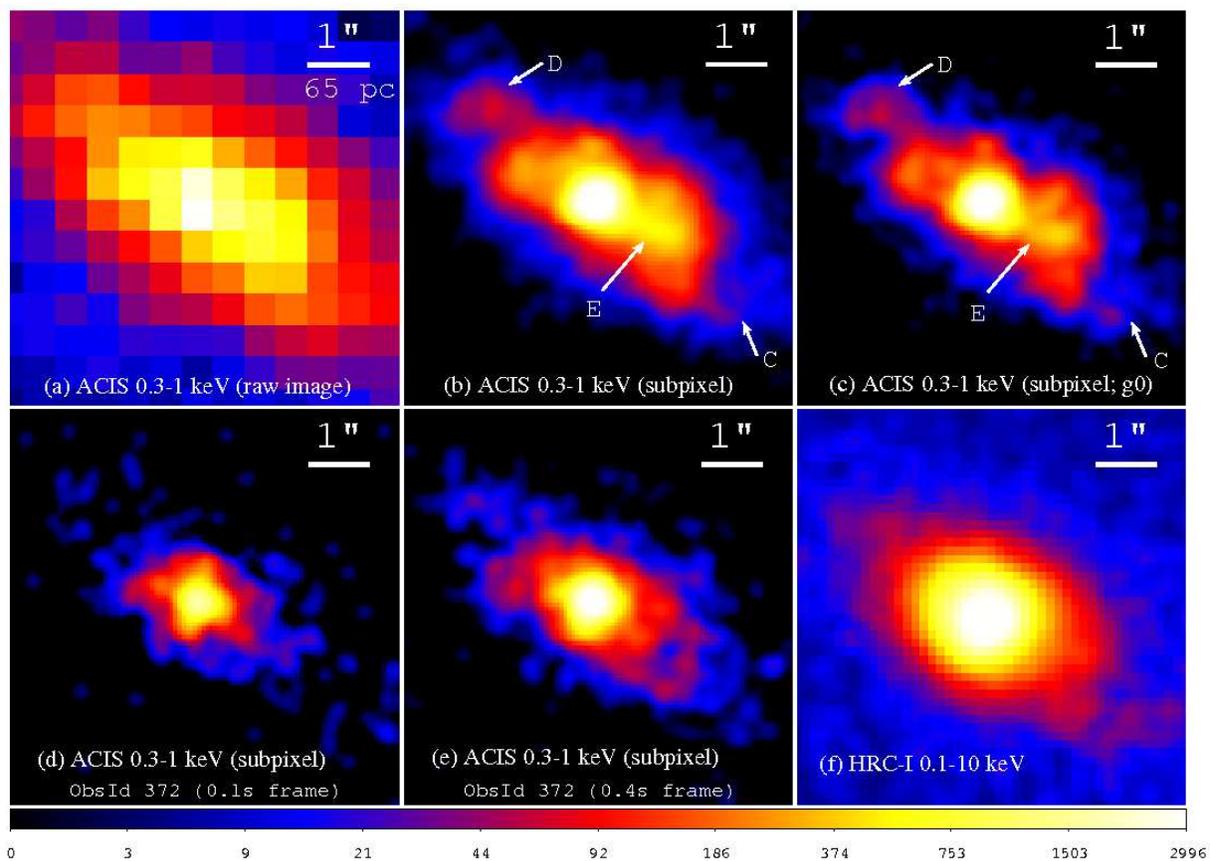} \caption{(a) Raw 0.3-1 keV ACIS image
  of the inner $6\arcsec \times 6\arcsec$ NGC 4151 nucleus; (b) ACIS
  image of the same region with subpixel binning (1/8 native pixel),
  demonstrating the improved resolution; (c) Same as (b) but using
  only single pixel events (grade 0); (d) 0.3-1 keV ACIS image (0.1 s
  frame time) from ObsID 372 with subpixel binning; (e) same as (d)
  but for a frame time of 0.4 s; (f) HRC image (0.1-10 keV). All
  subpixel resolution image were smoothed with a $FWHM=0.25\arcsec$
  Gaussian kernel for better visualization of faint extended features.
\label{fig4}}
\end{figure}

Figures~\ref{fig4}a and b zoom in on the inner $\sim 6\arcsec \times
6\arcsec$ (390~pc$\times$390~pc) nuclear region, with and without
subpixel resolution. The X-ray emission on scales of 200 pc ($\approx
3\arcsec$) is clearly extended along the NE-SW direction.  This extent
is in good agreement with the overall morphology of the [OIII]
emission (e.g. Kaiser et al.\ 2000; Wang et al.\ 2009). The ACIS
images with subpixel binning (e.g., Figures~\ref{fig3}a
and~\ref{fig4}b) reveal curvy extensions 2\arcsec\/ away from the
nucleus (labelled as ``C'' and ``D''), which are not discernable in
the raw soft ACIS image (Figure~\ref{fig4}a).

The subpixel ACIS image also shows a linear feature at
P.A.=235$^{\circ}$ (labelled as ``E'' in Figure~\ref{fig4}b) and
discrete knots in the inner region (on scales $<$1\arcsec\/ from the
nucleus).  The soft X-ray morphology is similar to that of the NLR
clouds seen in {\em HST} optical images of NGC 4151 (e.g., Evans et
al. 1993, Winge et al. 1999; see Wang et al. 2009 for a {\em
  Chandra}/HRC and {\em HST}/WFPC2 image comparison).  Corresponding
outer clouds were previously detected in deep {\em HST} images
(Hutchings et al. 1998; Kaiser et al. 2000).  Comparison of these
features to the [OIII] emission line clouds and radio outflows (e.g.,
Pedlar et al. 1992; Mundell et al. 2003), and emission modeling will
be presented in Paper II.

We emphasize that although the inner 2\arcsec\/ region is affected by
pile-up (\S~\ref{pile-up}) in the 0.3--1 keV band, the ACIS features
seen in the soft X-ray emission are not observation specific or
pile-up artifacts, for the following reasons:

\begin{enumerate}

\item For the on-axis nucleus, the PSF with pile-up simulated with
  MARX is circular, thus the elongated features (4\arcsec\/ across)
  seen in the soft band cannot be due to pile-up related PSF
  distortion;

\item Using only the single pixel events (grade 0), which give the
  best spatial resolution and are likely to be unaffected by pileup
  (Ballet 1999; Davis 2001), the extended features are clearly present
  (Figure~\ref{fig4}c);

\item Low pile-up ACIS images from {\em Chandra} archival observations
  of NGC 4151 (ObsID 372) show structures of similar shape and spatial
  scale as the features seen in our images (Figures~\ref{fig4}d and
  e).  Because of the short frame time (0.1 s and 0.4 s) and a factor
  of two lower nuclear flux, the estimated pile-up fraction is 6\% and
  11\% in these observations, respectively \citep{Yang01,Omai08}.  We
  also examined the HETG zero-order images of NGC 4151.  Although some
  suffer significant pile-up due to longer frame time and a brightened
  nucleus (ObsID 7830), extended emission along NE--SW with similar
  substructures as we detect is seen in ObsIDs 335 and 7829;

\item The PSF-deconvolved HRC image (Figure~2 in Wang et al. 2009),
  which is not affected by pile-up, shows the same extended morphology
  and small-scale enhancements within 2\arcsec\/ from the nucleus that
  are not seen in ACIS without subpixel binning.

\end{enumerate}

\begin{figure}
\epsscale{1.0} \plotone{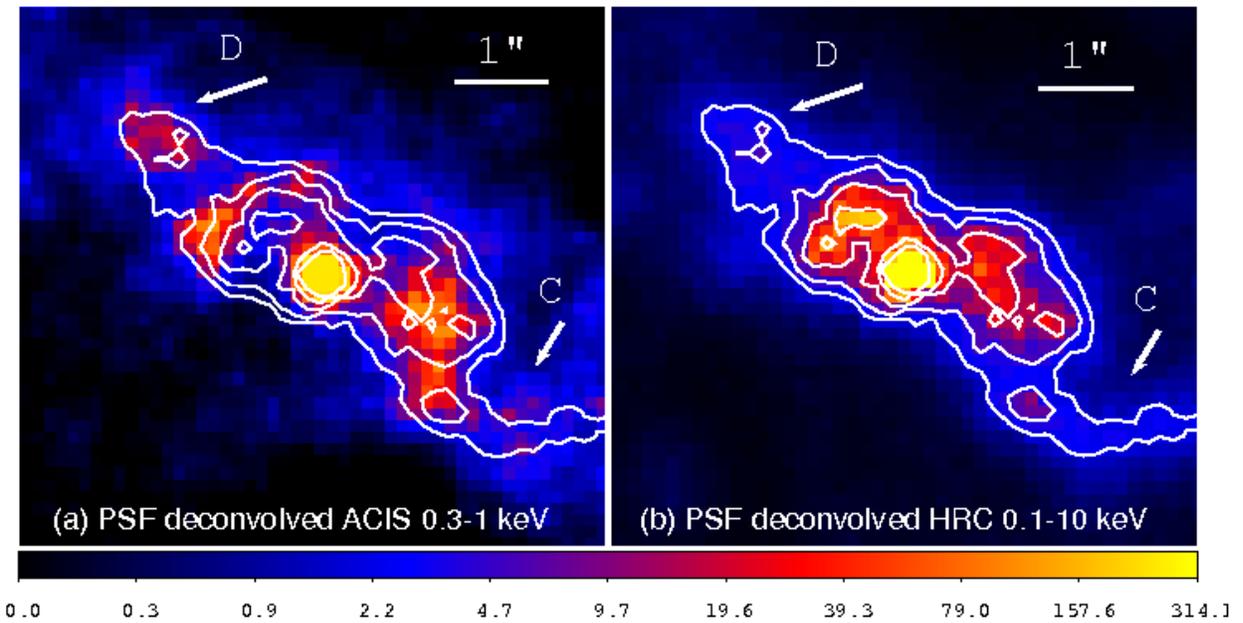} \caption{(a) PSF-deconvolved ACIS
  image (0.3--1 keV) using the EMC2 method (100 iterations; Esch et
  al. 2004; Karovska et al.\ 2007); (b) PSF-deconvolved HRC image
  (0.1--10 keV; Wang et al.\ 2009) using the EMC2 method. Contours of
  the deconvolved HRC image shown in (b) are overlaid in both panels
  to assist comparison of the features.\label{fig5}}
\end{figure}

We applied two image reconstruction techniques to the ACIS data, the
Richardson-Lucy (R-L) algorithm (Richardson 1972; Lucy 1974) and the
Expectation through Markov Chain Monte Carlo (EMC2) method (Esch et
al.\ 2004; Karovska et al.\ 2007). Both show similar extended
morphology, although the R-L deconvolution produces more grainy
restoration of extended features.  Figure~\ref{fig5} compares the EMC2
PSF deconvolved ACIS image with the similarly deconvolved HRC image
from Wang et al.\ (2009).  The resemblance of the NE-SW elongation and
substructures (such as the curvy feature ``C'' and NE clump ``D''
marked in Figure~\ref{fig3}a) between the subpixel ACIS images and the
deconvolved HRC strengthens our confidence that the resolved features
are real.  However, we noticed that the X-ray clump 1\arcsec\/ NE of
the nucleus in the ACIS image is displaced in the HRC image
($\sim$0.3\arcsec\/ closer to the nucleus).  The morphology of the
enhancement located 1\arcsec\/ SW to the nucleus is also different in
two reconstructed images: it is mostly a linear north-south elongation
in the ACIS image, whereas in the HRC image it shows a tail towards
SW).  Two factors may cause these discrepancies: (1) The region within
1\arcsec\-radius of the nucleus in the ACIS image is affected by
pile-up, but the accuracy of the piled PSF on this scale has not been
calibrated; yet this inaccuracy may have an impact on the
reconstructed image; (2) HRC is sensitive to photons in the energy
range of 0.1--10 keV, therefore may reveal real structures that are
not seen in the 0.3--1 keV ACIS image.

\subsection{Hardness Ratio Image of the Central 3 kpc-diameter Region}

\begin{figure}
\epsscale{0.5} \plotone{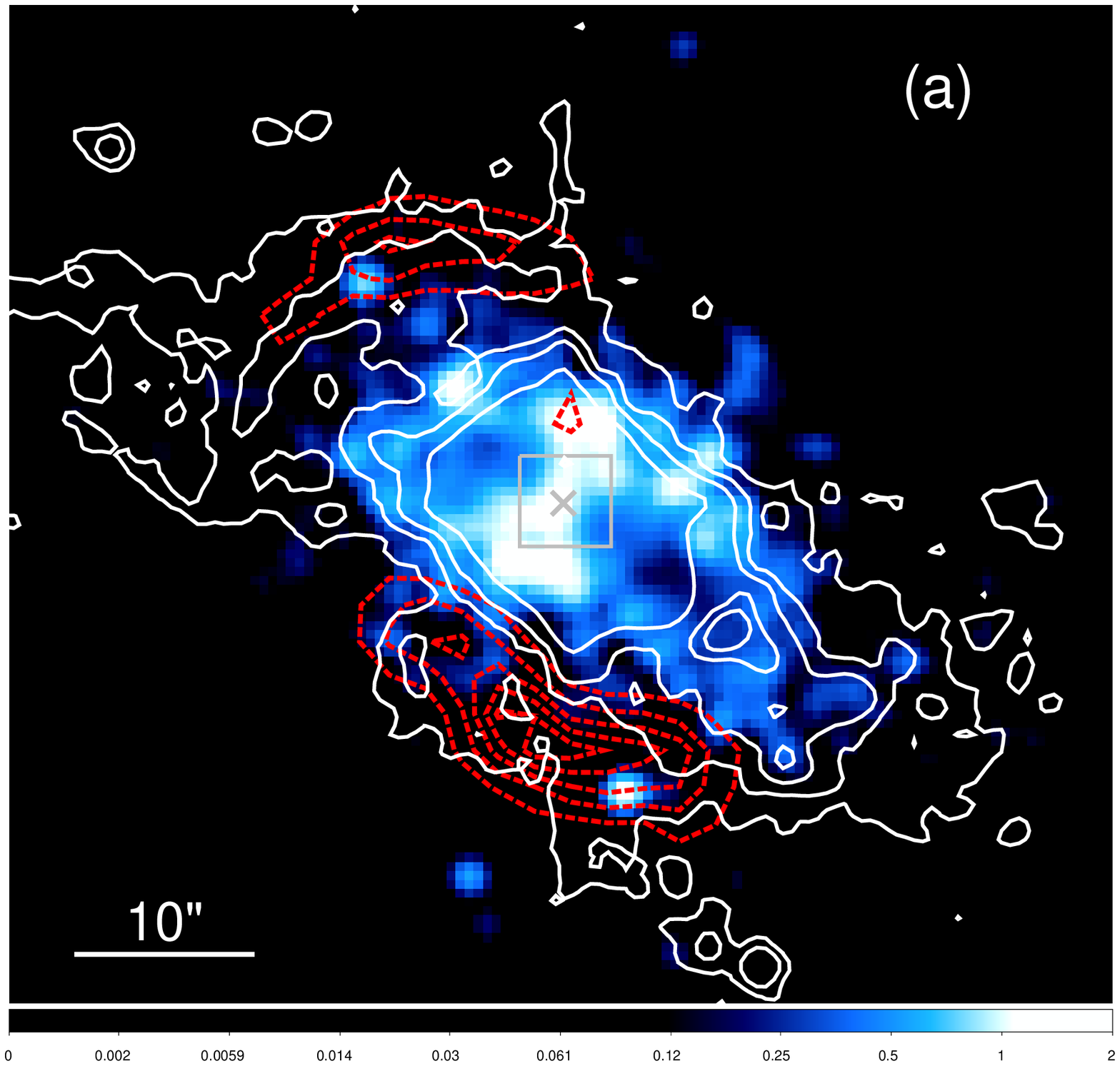} \plotone{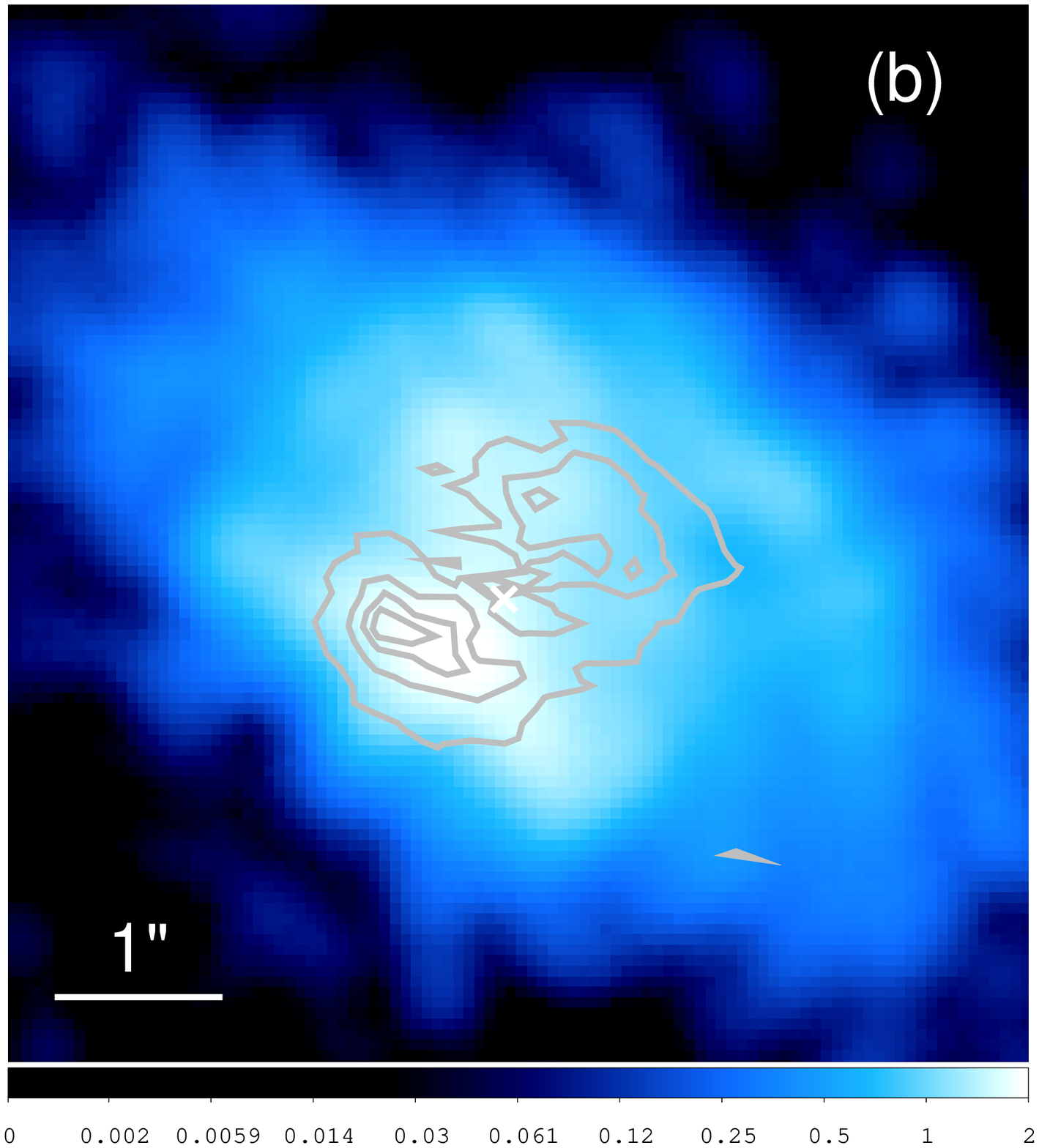}
\caption{(a) Hardness ratio map of the central $1\arcmin \times
  1\arcmin$ (3.9~kpc-across) region, defined as the ratio of the
  counts in the 1--2 keV to the 0.3--1.0 keV band. Contours in white
  (solid) and red (dashed) show the continuum-subtracted H$\alpha$
  \citep{Knapen04} and $^{12}$CO emission \citep{Dumas10},
  respectively. (b) Same as (a) but zoomed-in to the inner $6\arcsec$
  region outlined by a box in (a). The cross marks the location of the
  nucleus. Grey contours show the H$_2$ $\lambda$2.1218~$\mu$m
  emission \citep{SB09}. \label{fig6}}
\end{figure}

To find regions of harder spectral index or higher X-ray obscuration,
we extracted {\em Chandra} images in the 0.3--1.0 keV and 1.0--2.0 keV
bands and derived the counts ratio $C_{1-2 {\rm keV}}$ over $C_{0.3-1
  {\rm keV}}$.  These bands are sensitive to hydrogen column densities
of the order of $10^{22}$ cm$^{-2}$, and the relative changes in the
ratio map have been demonstrated to reflect variations in the X-ray
absorption (Bianchi et al. 2007).  The resulting hardness ratio map of
the central $50\arcsec$ (3.2 kpc) diameter region is shown in
Figure~\ref{fig6}a, with H$\alpha$ \citep{Knapen04} and CO
\citep{Dumas10} contours outlining the ionized gas and the cold
molecular gas lanes, respectively.  Figure~\ref{fig6}b shows the
hardness ratio map of the inner 6\arcsec\/ (400 pc) diameter region.

One obvious feature in Figure~\ref{fig6}a is the harder (white) region
extended in the NW--SE direction (P.A. $\sim$150$^{\circ}$).  This is
approximately the P.A. of the minor axis of the galaxy and the weak
bar (Mundell \& Shone 1999), and perhaps also the direction of a
putative nuclear torus, mapped in H$_2$ emission by \citet{SB09}
(contours in Figure~\ref{fig6}b; see also Fernandez et
al.\ 1999). Thus higher obscuration is plausible in this region.
Coincident with the higher hardness ratio region across the nucleus
(Figure~\ref{fig6}b) is a dark dust patch in the $V-I$ image
\citep{Asif98,Ohtani01}, noted by \citet{Terlevich91} in their $V-R$
image as ``the red bar'' (P.A.=150$^{\circ}$; 1\arcsec\/ in length).

\begin{figure}
\epsscale{0.6} \plotone{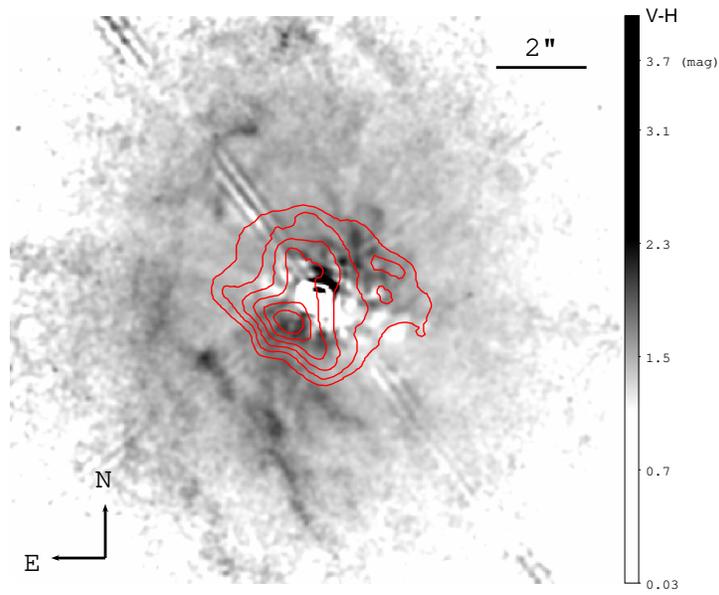}
\caption{$V-H$ color map of the NGC 4151 nuclear region from {\em HST}
  F550M and F160W band images. Point source has been removed using the
  simulated PSF and the resolution of F550M image was degraded to
  match that of the F160W image. Contours (red) are from the X-ray
  hardness ratio map shown in Figure~\ref{fig6}b.  Note that residuals
  from diffraction spikes can still be seen at
  P.A.$\approx$40$^{\circ}$/220$^{\circ}$.\label{fig7}}
\end{figure}

To examine the extinction of the nuclear region of NGC 4151, we
created the $V-H$ color map using {\em HST} images (\S~\ref{obs_2}),
covering the central $10\arcsec \times 10 \arcsec$ of NGC 4151
(Figure~\ref{fig7}).  Dark spiral features of higher extinction are
clearly revealed towards the nucleus perpendicular to the bicone
direction, thanks to the high spatial resolution afforded by {\em
  HST}.  Adopting the extinction curve in \citet{CCM89}, the observed
$E(V-H)$ suggests a range between 0.5 and 1 for $E(B-V)$, which is
equivalent to $1.5\la A_V \la 3$ (mag).

Another notable feature in Figure~\ref{fig6}a is that the edges of the
ENLR show harder emission than the inner region.  To obtain
quantitative constraints on the hardness ratios, we calculated the
uncertainties using the Bayesian estimation of Hardness Ratios (BEHR;
Park et al. 2006) tool provided by the California-Harvard
Astro-Statistics
Collaboration\footnote{\url{http://hea-www.harvard.edu/AstroStat/}},
where the hardness ratio, $HR$, is defined in BEHR as $(C_{1-2 {\rm
    keV}}-C_{0.3-1 {\rm keV}})/(C_{1-2 {\rm keV}}+C_{0.3-1 {\rm
    keV}})$.  For the inner part of the cone (named R1)
$HR=-0.63^{+0.01}_{-0.02}$, for the edges of the optical outflow (R2)
$HR=-0.27\pm 0.03$, and for the highest hardness ratio region
perpendicular to the cone-axis (R3) $HR=-0.04\pm 0.02$.  Thus the
hardness ratios differ significantly for each of these regions.  This
is consistent with the hollow bi-cone geometry proposed in Das et
al.\ (2005) (see also Evans et al. 1993; Robinson et al. 1994;
Hutchings et al. 1998; Storchi-Bergmann et al. 2010), where clumpy
outflowing material is mostly located along the boundaries of the
bi-cone.  One side of the cone is expected to intersect an area of the
host galactic disk, and the projected appearance will be an
illuminated narrow sector (e.g., region R1) with higher absorption
along the edges, which is indeed observed here (region R2).

To further investigate these absorption column density variations,
spectra from the three regions above were extracted using the hardness
ratio map, as outlined in Figure~\ref{fig8}.  Because the PSF
scattered emission is not negligible for the bright nucleus, we fit
the spectra with a simple model consisting of an absorbed soft thermal
component and a nuclear component (Wang et al. 2010b), allowing the
normalization of the nuclear component to vary to account for
contribution from the PSF wings.  Figure~\ref{fig9}a-c show the
spectra and the fits; the fitting results are summarized in
Table~\ref{tabspec}.  These spectra confirm the wide range of X-ray
absorption suggested by the hardness ratio map.  The edges of the cone
(R2) indeed show a higher absorbing column ($N_H=7\pm 5\times 10^{21}$
cm$^{-2}$) than the inner cone region (R1), for which only a Galactic
column ($N_H=2\times 10^{20}$ cm$^{-2}$; Murphy et al. 1996) is
required.  The highest column density, $N_H=6.5\pm 1.5 \times 10^{22}$
cm$^{-2}$, is found across the nucleus (R3) and is approximately
equivalent to $E(B-V)=10$ adopting $N(HI+H_2)/E(B-V)=5.8\times
10^{21}$ cm$^{-2}$ mag$^{-1}$ \citep{Bohlin78}.  This value is
significantly higher than the average $E(B-V)\approx 0.5$ found in the
NLR \citep{SB09} and the $E(B-V)=0.5-1.0$ found in the $V-H$ map.
\citet{Mundell95} find an HI absorption column density of $3.9\times
10^{21}$ cm$^{-2}$ towards the radio nucleus; the factor $\sim$15
higher X-ray $N_H$ suggests that most of the intervening hydrogen is
in the form of $H_2$ or high abundances $\sim 15 Z/Z_{\odot}$.

\section{Discussion}

\subsection{Extended X-ray Morphology}\label{discuss1}

This is the first time such a high sensitivity high resolution X-ray
spectral imaging of the NGC 4151 nuclear region has been obtained. The
HRC image provides no spectral information, and previous ACIS images
(Ogle et al. 2000; Yang et al. 2001) either lack sensitivity to
convincingly identify the X-ray features seen in subpixel resolution,
or suffer from heavy pile-up in the inner 3\arcsec\/.

The morphological features of the X-ray emission of NGC 4151 are
summarized as follows: (1) By pushing the spatial resolution of the
ACIS image with subpixel resolution techniques, X-ray enhancements in
the innermost nuclear region ($r\la 1\arcsec$) seen in the
PSF-deconvolved HRC image are recovered; (2) The extended soft X-ray
emission shows the same biconical morphology as the optical ENLR out
to a radius (20\arcsec\/; $\sim$1.3 kpc) larger than seen in earlier
X-ray studies (c.f. 14\arcsec; $\sim$900 pc, Yang et al. 2001); (3)
There is also firm detection of extended X-ray emission at angles
beyond the boundaries of the optical ENLR ionization cone; (4) The
observed X-ray emission is more absorbed towards the edge of the cone
and perpendicular to the cone axis at the apex of the cone.

The two-sided, bifurcated X-ray emission morphology is consistent with
the model described in Das et al. (2005) and \citet{SB10}, in which
clouds are outflowing along the walls of a hollow bicone inclined
$45^{\circ}$ to the line of sight.  One side of the cone is expected
to exit the plane at 21$^{\circ}$, intersecting an elongated area of
the host galactic disk.  The projected appearance \citep[c.f. Figure 9
  in][]{SB10} closely resembles the observed morphology here.

The extended X-ray emission seen perpendicular to the ENLR, along the
NW-SE direction, was first suggested to be ``unresolved or marginally
resolved'' in Yang et al. (2001).  The definitive resolved emission at
angles beyond the edges of the bicone (Figure~\ref{fig2}c) indicates
that the gas there is not completely shielded from the continuum, as
would be the case for a molecular torus collimating the bicone.  This
agrees well with the detection of significant optical emission outside
the emission-line bicone in the {\em HST} [OIII] and [OII] images,
where Kraemer et al. (2008) were able to model the emission lines with
a weaker ionizing continuum, filtered by an ionized absorber partially
covering the nuclear continuum, i.e. the accretion disk wind
\citep[e.g.,][]{Elvis00}.

The NE soft X-ray emission reaches as far as 11\arcsec\/ (700 pc) from
the nucleus, then the surface brightness decreases rapidly and
significantly, by a factor of 5.  This location lies near to both the
northern CO gas lane and one of the dust arcs identified in the
optical $V-I$ color map \citep[a region not covered by our $V-H$
  map]{Asif98}. These features are gaseous compressions in
circumnuclear gas spiral arms due to the gravitational influence of
the large-scale stellar bar \citep{Mundell99,Dumas10}.  Since the
biconical geometry \citep{Das05} implies that both cones are
intersecting the host galactic disk, the terminus of the NE X-ray
emission is likely to occur where the NE outflow encounters dense ISM
in the galactic disk.  We will discuss this boundary further in Paper
II.

\subsection{X-ray Absorption Features and Cold Material for AGN Feeding}

\begin{figure} 
\plotone{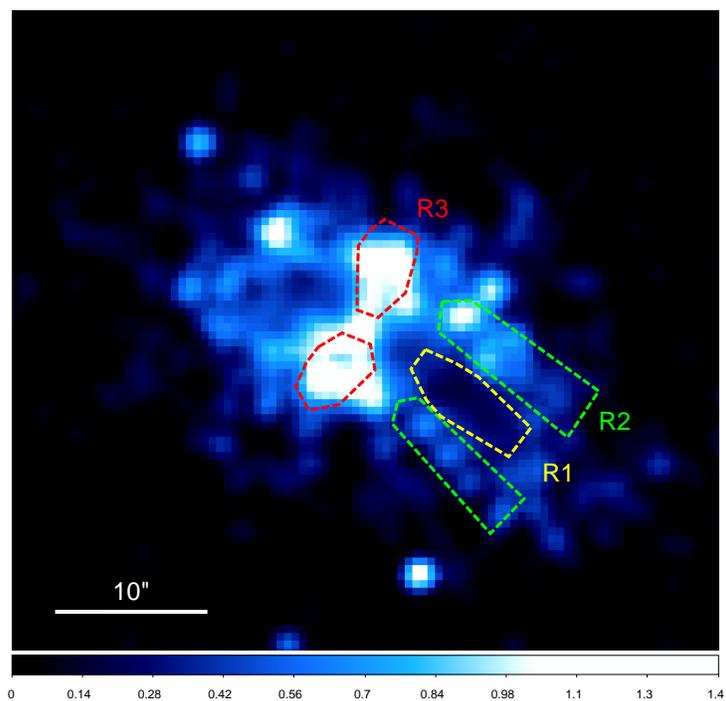}
\caption{The spectral extraction regions overlaid on the hardness
  ratio image. The polygons outline the regions R1 (inner cone
  region), R2 (edges of the cone), and R3 (region perpendicular to the
  bicone-axis). \label{fig8}}

\end{figure}

\begin{figure}
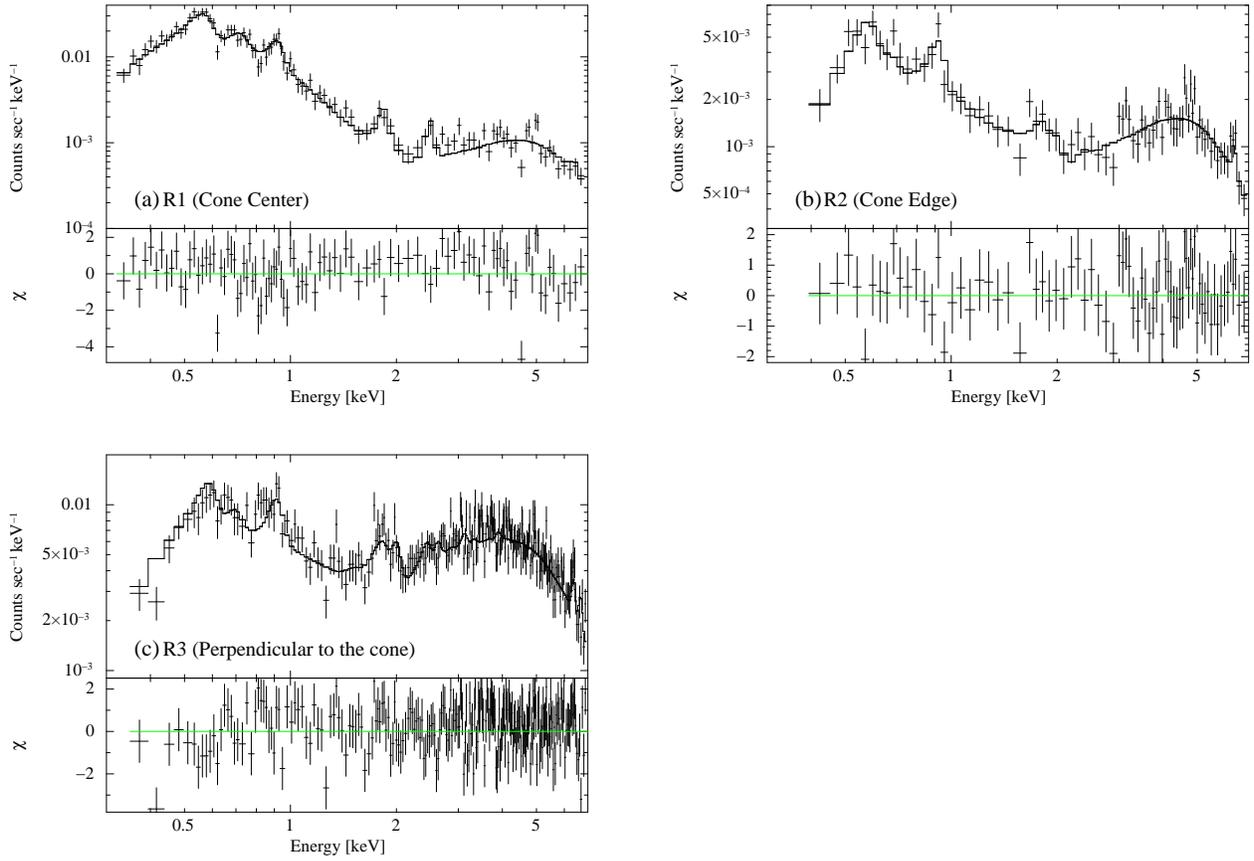
 
\includegraphics[scale=.30,angle=-90]{f9a.eps}
\includegraphics[scale=.30,angle=-90]{f9b.eps}
\includegraphics[scale=.30,angle=-90]{f9c.eps}
\caption{The X-ray spectra and fits obtained for three regions
  defined in Figure~\ref{fig9} that are morphologically distinct (see
  text and Table~\ref{tabspec}). \label{fig9}}

\end{figure}

The features seen in the hardness ratio map (Figure~\ref{fig6}a) also
agree well with the bicone models \citep{Das05,SB10}.  In the geometry
illustrated by \citet{Das05}, the more absorbed X-ray emission seen
along the bi-cone edges (R2 in Figure~\ref{fig8}) can be explained by
higher absorption from the clumpy medium outflowing along the wall of
the SW cone, as it emerges from the galactic disk.  Adopting the
conical geometry in Das et al. (2005) with an inner full opening angle
of 30$^{\circ}$ and outer full opening angle of 66$^{\circ}$, and
assuming that the $N_H$ is due to the extra optical path along the
cone wall, we estimate an average $n_H \sim 10$ cm$^{-3}$ in the cone
wall (c.f. $n_H=220$ cm$^{-3}$ for optical clouds; Penston et
al. 1990).  The region perpendicular to the bi-cone axis (R3 in
Figure~\ref{fig8}) may represent the inward extension of the CO gas
lanes.

The innermost $r\leq 1.5\arcsec$ region (P.A.$\sim$150$^{\circ}$) of
high X-ray hardness ratio (Figure~\ref{fig6}b) is spatially coincident
with the H$_2$ $\lambda 2.1218$ $\mu$m emission \citep[grey
  contours;][]{SB09,Fer99} and the dusty spirals in the $V-H$ color
map (Figure~\ref{fig7}), which may trace inflowing material feeding
the AGN.  Next we discuss three mechanisms that were considered by
\citet{SB09} for the origin of H$_2$ line emission: (1) fluorescence
by FUV photons; (2) thermal excitation by shocks in the accretion flow
towards the nucleus, or (3) photo-excitation by X-rays from the AGN.

(1) {\em Fluorescence by FUV photons} is ruled out by \citet{SB09}
based on the H$_2$ 2-1 S(1)/1-0 S(1) line ratio and strong evidence
for thermal equilibrium \citep[$T_{exc}=2155$~K;][]{SB09} in the H$_2$
lines.  However, it is worth noting that the peak density of
$10^4$--$10^5$ cm$^-3$ in the H2 emitting region, taken at the
measured value within the inner 0.5 arcsec in Storchi-Bergmann et
al. (2009), is above the critical density of the
rotational-vibrational transitions of H$_2$ ($\sim 10^4$ cm$^{-3}$;
Tielens 2005).  It implies that thermalization by collisions
effectively erases the signature of fluorescent excitation (Black \&
van Dishoeck 1987; Draine \& Bertoldi 1996; Hollenbach \& Tielens
1999), therefore the line ratio diagnostics used at lower densities
may become uncertain to distinguish whether UV fluorescence or
collisional excitation is dominating the H$_2$ emission.

(2) {\em H$_2$ excitation by shocks in the accretion flow} is not
favored by current data. Mundell \& Shone (1999) interpreted the H$_2$
emission as being associated with the molecular torus obscuring the
AGN which would explain its orientation perpendicular to the radio
jet.  The inflows observed in HI observations (Mundell \& Shone 1999)
are only traced to the inner part of the oval bar ($r\approx
20\arcsec$; Asif et al. 1998).  On the 1 kpc scale, most of the cold
molecular gas is distributed along the curved CO lanes
$\sim$15\arcsec\/ away from the nucleus (Dumas et al. 2010) with
kinematics consistent with the disk and the bar (e.g., Mundell et
al. 1999).  No strong kinematic shocks are expected, which is also
consistent with the gravitational potential that shows no important
non-axisymmetric component on the relevant scales.  Moreover, the
detection of H$_2$ in this relatively high density region ($n_H\sim
10^5$ cm$^{-3}$) is evidence against the presence of relatively strong
shocks ($v_s \ga 25$ km s$^{-1}$) in this region, which would
dissociate the molecular hydrogen collisionally.  However, it remains
possible that weaker shock flows exist and contribute to the H$_2$
excitation.

(3) {\em Photo-excitation of H$_2$ by the AGN} can be evaluated more
quantitatively following \citet{Riffel08} and \citet{SB09}, using the
line intensities calculated for X-ray-irradiated molecular gas by
\citet{Maloney96} and our measured parameters.  A key parameter is the
effective ionization parameter, $\xi_{\rm{eff}}=1.26\times 10^{-4}
F_x/(n_5 N_{22}^{0.9})$ \citep[Equation 12 in][]{Maloney96}, where
$F_x$ is the incident X-ray flux, $n_5$ is the cloud density in units
of $10^5$ cm$^{-3}$ and $N_{22}$ is the absorption column in units of
$10^{22}$ cm$^{-2}$.  For $F_x=7.6$ erg cm$^{-2}$ s$^{-1}$ ($L_{{\rm
    2-10 keV}}=1\times 10^{42}$ erg s$^{-1}$; Wang et al. 2010b) at
the H$_2$ emission peak \citep[$\sim$30~pc from the nucleus;][]{SB09},
$n_5=1$, and $N_{22}=6.5$ (Table~2), we obtain $\xi_{\rm{eff}}\sim
2\times 10^{-4}$ and an emerging H$_2$ 1-0 S(1) intensity of $\sim
8\times 10^{-5}$ erg cm$^{-2}$ s$^{-1}$ sr$^{-1}$.  For the aperture
($0.3\times 0.3$ arcsec$^2$ or $2.16\times 10^{-12}$ sr) used by
\citet{SB09}, the expected value is $2\times 10^{-16}$ erg cm$^{-2}$
s$^{-1}$.  This is a factor of two higher, but likely consistent with,
the observed line flux for the same transition \citep[$1\times
  10^{-16}$ erg cm$^{-2}$ s$^{-1}$;][]{SB09}, given the uncertainties
in our assumptions and the models.  Thus, excitation by the X-rays
from AGN can reasonably explain the observed H$_2$ line emission, in
agreement with the conclusion drawn in \citet{SB09}.  This also
indicates the X-ray radiation received by the H$_2$ gas is comparable
to that observed along our line-of-sight, which is not blocked by a
molecular torus in this direction.

The X-rays from the AGN may also explain the non-detection of CO gas
in contrast to the $10^7-10^9 M_{\odot}$ molecular gas estimated by
\citet{SB10}. Dumas et al. (2010) showed that no CO gas was detected
within the inner 300 pc of the active nucleus, where strong H$_2$ line
emission arises.  Using a standard Galactic CO-to-H$_2$ conversion
factor of $X_{\rm CO} = 2\times 10^{20}$ cm$^{-2}$ (K km
s$^{-1}$)$^{-1}$ (Solomon \& Barrett 1991), \citet{Dumas10} derived a
1-$\sigma$ upper limit of $10^5 M_{\odot}$ for the cold molecular gas
mass.  They suggested that CO itself may not be a good tracer of
molecular gas in X-ray dominated regions (XDR) close to an active
nucleus, where equilibrium molecular abundances are affected
(Meijerink \& Spaans 2005, 2007).

Model 3 in Meijerink \& Spaans (2005) calculated the CO/H$_2$
abundance ratio for $F_x=1.6$ erg cm$^{-2}$ s$^{-1}$ and $n=10^{5.5}$
cm$^{-3}$ (the model parameters most close to the flux and density
seen here), which is $\ll 10^{-4}$ at column densities below
$N_H=10^{23}$ cm$^{-2}$.  For clouds closer to the nucleus or immersed
in diffuse X-ray emission, $F_x$ can be higher, and a CO/H$_2$ ratio
of $>10^{-4}$ can only be found at column densities $N_H \ga 10^{24}$
cm$^{-2}$ (model 4 in Meijerink \& Spaans 2005).  Given that the
absorption column from the spectral fit is $N_H\sim 6.5\times 10^{22}$
cm$^{-2}$, the lack of CO perhaps can be explained by a low CO
abundance relative to H$_2$ in the XDR.

Assuming that most intervening material is in the form of H$_2$, the
X-ray derived $N_H=6.5\pm1.5 \times 10^{22}$ indicates a total H$_2$
mass of $2\times 10^7$~M$_{\odot}$ in the $3\arcsec \times 3\arcsec$
(200~pc diameter) nuclear region imaged by \citet{SB09}.  This value
agrees well with the lower end of total H$_2$ mass
($10^7-10^9$~M$_{\odot}$) in \citet{SB09}, which was derived using the
mass of ``hot'' H$_2$ (240~M$_{\odot}$) detected in the near-IR.

One clear prediction of the XDR scenario is that, like in shocks, the
[OI] 63$\mu$m/[CII] 158$\mu$m ratio is higher ($\ga 10$) than
generally found in PDRs \citep{Maloney96}.  This can be tested with
new observations using far-infrared (FIR) facilities such as PACS
spectrometer (Poglitsch et al. 2010) on board the {\em Herschel Space
  Observatory} (Pilbratt et al. 2010).  NGC 4151 was observed as part
of {\em Herschel} Key Program SHINING (PI: Sturm).  They find that the
[O I] 63$\mu$m emission is highly concentrated on the central spaxel,
while the [C II] 158$\mu$m emission is more extended.  The ratio of [O
  I] 63$\mu$m / [C II] 158$\mu$m in the central spaxel is high,
$\sim$11 (E. Sturm, private communication), consistent with XDR or
shock models with an effective ionization parameter of $\log \xi_{\rm
  eff} \sim -3$ (Figure 12 in Maloney et al.\ 1996).  We note that
this value is higher than $\log \xi_{\rm eff}= -3.7$ in the XDR model
analysis above using the observed X-ray flux and column density.  This
discrepancy might be attributed to more than one mechanism at work
exciting the H$_2$ emission.  Similar [OI]/[CII] ratios of $\sim$10
have been reported in dense PDRs with no significant X-ray excitation,
for example in regions near the ionizing stars of the Orion Nebula
(Boreiko \& Betz 1996).

In short, more investigations on the importance of UV fluorescence,
shock excitation, and X-ray excitation of H$_2$ emission in NGC 4151
are needed to further examine the caveats discussed above.  In
particular, theoretical effort on models that focus on the
astrophysical environment around AGN will be valuable to clarify the
ambiguity of the H$_2$ excitation mechanism in NGC 4151 and in other
systems in general.

\section{Conclusions}

We have obtained and analyzed deep {\em Chandra} observation of NGC
4151, which demonstrated the imaging power of ACIS with subpixel
resolution to resolve the circumnuclear X-ray emission of a nearby
Seyfert galaxy.

The X-ray structures in the nuclear region (inner 300 pc) seen in our
HRC study of NGC 4151 (Wang et al. 2009) are recovered in the deep
ACIS image with subpixel resolution techniques.  Moreover, the
extended soft X-ray emission of NGC 4151 reaches to a distance of
$\sim$1.3 kpc to the SW of the nucleus, much farther than seen in
earlier X-ray studies.  We suggest that the terminus of the NE X-ray
emission is likely where the NE outflow encounters dense ISM in the
galactic disk.  Extended X-ray emission is firmly detected outside the
boundaries of the ionization cone, which indicates that the continuum
is not completely blocked by a molecular torus.  This is consistent
with the findings of optical studies where nuclear ionizing photons
were filtered by the bi-cone wind.

The hardness ratio image and spectral fits show that the X-ray
emission is more absorbed towards the edges of the cone, as well as
perpendicular to the cone axis along the P.A. of the putative nuclear
torus.  These regions of different spectral hardness are consistent
with the expected features in the biconical outflow models.

X-ray emission shows the highest hardness ratio in the innermost
region, spatially coincident with the near-infrared resolved H$_2$
emission and the dusty spirals revealed in the {\em HST} ($V-H$) color
map. All three features possibly trace the cold material feeding the
central SMBH.  The agreement between the observed H$_2$ line flux and
the predicted value from X-ray-irradiated molecular cloud models
supports photo-excitation by the X-ray radiation from the active
nucleus, which is not blocked by a molecular torus in this direction.
However, with current data, the contribution from UV fluorescence or
collisional excitation cannot be ruled out.

The X-rays from the AGN also explain the recent non-detection of CO
gas \citep{Dumas10} in contrast to the presence of $10^7-10^9
M_{\odot}$ molecular hydrogen gas \citep{SB09}. The total H$_2$ mass
derived from the X-ray observation agrees with that expected from the
mass of ``hot'' H$_2$.  The absorption column derived from the X-ray
spectral fit, when compared to the XDR models, implies that the lack
of CO close to the nucleus may be explained by its low abundance
relative to H$_2$ in the XDR.

\acknowledgments

We thank the anonymous referee for helpful comments on the H$_2$
excitation mechanisms.  This work is supported by NASA grant
GO8-9101X.  We acknowledge support from the CXC, which is operated by
the Smithsonian Astrophysical Observatory (SAO) for and on behalf of
NASA under Contract NAS8-03060.  CGM acknowledges financial support
from the Royal Society and Research Councils U.K. GD was supported by
DFG grants SCH 536/4-1 and SCH 536/4-2 as part of SPP 1177.  We thank
Eckhard Sturm for providing the FIR line ratios prior to publication
of their work, Dan Harris and Dan Schwartz for advice on subpixel
binning.  J. W. thanks T. Storchi-Bergmann and R. Riffel for providing
the Gemini NIFS maps, M. McCollough, E. Galle, M. Juda and T. Gaetz
(CXC) for technical discussion on {\em Chandra} PSF, ACIS readout
streaks and HRC data reduction.  This research has made use of data
obtained from the {\em Chandra} Data Archive, and software provided by
the CXC in the application packages CIAO and Sherpa.  This research
used observations made with the NASA/ESA Hubble Space Telescope, and
obtained from the Hubble Legacy Archive, which is a collaboration
between the Space Telescope Science Institute (STScI/NASA), the Space
Telescope European Coordinating Facility (ST-ECF/ESA) and the Canadian
Astronomy Data Centre (CADC/NRC/CSA).

{\it Facilities:} \facility{CXO (HRC, ACIS)}

\clearpage

\clearpage
\clearpage

\clearpage

\clearpage

\clearpage

\clearpage

\clearpage

\clearpage

\clearpage

\begin{deluxetable}{lcccc}
\tabletypesize{\scriptsize} \tablecaption{Log of {\it Chandra} ACIS-S Observations of NGC 4151\label{obslog}} \tablewidth{0pt}

\tablehead{\colhead{Obs ID}& \colhead{Date} & \colhead{Instrument} & \colhead{Frame Time (s)}  & \colhead{Exposure (ks)}}
\startdata
372 & Mar 06, 2000 & ACIS-S  & 0.1 & 0.8\\
372 & Mar 06, 2000 & ACIS-S  & 0.4 & 6.2\\
9217& Mar 29, 2008& ACIS-S& 0.6 &116 \\
9218& Mar 27, 2008& ACIS-S& 0.6 & 63 \\
9219& Mar 02, 2008& HRC-I & $1.6\times 10^{-5}$ & 49 \\
\enddata
\end{deluxetable}

\clearpage \pagestyle{empty}
\begin{deluxetable}{cccccc}
\tabletypesize{\scriptsize}
\tablewidth{0pt} \tablecaption{Spectral Fits to Extended
X-ray Emission Features \label{tabspec}} \tablecolumns{6}
\tablehead{\colhead{Region\tablenotemark{a}} & \colhead{$HR$\tablenotemark{b}} & \colhead{$N_H$} &
\colhead{Temperature} & PSF fraction\tablenotemark{c}  & \colhead{$\chi^2$ / d.o.f.}\\
\colhead{} & \colhead{} & \colhead{[$\times 10^{22}$ cm$^{-2}$]} & \colhead{[keV]} & \colhead{[$\%$]} & \colhead{}}

\startdata
R1 & $-0.63^{+0.01}_{-0.02}$& 0.02\tablenotemark{d} & 0.28$^{+0.02}_{-0.01}$ & 0.5 & 135/94 \\
R2 & $-0.27\pm 0.03$ & 0.70$^{+0.45}_{-0.54}$ & 0.08$^{+0.09}_{-0.05}$ & 0.7  & 66/75 \\
R3 & $-0.04\pm 0.02$ & 6.5$\pm$1.5 & $1.94^{+0.48}_{-0.51}$ & 2.2 & 261/231 \\
\enddata

\tablenotetext{a}{The regions and spectra are presented in
  Figure~\ref{fig9}. R1--the central inner part of the cone; R2--the
  edges of the cone; and R3--the high hardness ratio region
  perpendicular to the bicone-axis.} \tablenotetext{b}{Hardness ratio
  is defined as $(C_{1-2 {\rm keV}}-C_{0.3-1 {\rm keV}})/(C_{1-2 {\rm
      keV}}+C_{0.3-1 {\rm keV}})$.}  \tablenotetext{c}{The fraction of nuclear
  emission contributed to the extended emission.}
\tablenotetext{d}{Absorption fixed at Galactic column as required by
  the fit.}

\end{deluxetable}

\end{document}